\documentclass{article}
\usepackage{spconf,amsmath,graphicx}
\usepackage{amsfonts}
\usepackage{booktabs}
\usepackage{microtype}
\usepackage{hyperref}

\title{End-to-end real time tracking of children's reading with pointer network}
\name{Vishal Sunder, Beulah Karrolla, Eric Fosler-Lussier}
\address{The Ohio State University}
\begin{document}
%
\maketitle
\begin{abstract}
In this work, we explore how a real time reading tracker can be built efficiently for children's voices. While previously proposed reading trackers focused on ASR-based cascaded approaches, we propose a fully end-to-end model making it less prone to lags in voice tracking. We employ a pointer network that  directly learns to predict positions in the ground truth text conditioned on the streaming speech. To train this pointer network, we generate ground truth training signals by using forced alignment between the read speech and the text being read on the training set. Exploring different forced alignment models, we find a neural attention based model is at least as close in alignment accuracy to the Montreal Forced Aligner, but surprisingly is a better training signal for the pointer network. Our results are reported on one adult speech data (TIMIT) and two children's speech datasets (CMU Kids and Reading Races). Our best model can accurately track adult speech with 87.8\% accuracy and the much harder and disfluent children's speech with 77.1\% accuracy on CMU Kids data and a 65.3\% accuracy on the Reading Races dataset. 
\end{abstract}
\section{Introduction}
\label{sec:intro}

Tracking read speech finds useful applications in education, when teaching children how to read properly. Building reading tutors has been a popular application of automatic speech recognition (ASR) \cite{adams2013promise, mostow1994prototype, mostow2012and} and tracking is an important part of that. In contrast to offline assessment to score pronunciations and give offline feedback \cite{black2007automatic, wu2021transformer, venkatasubramaniam2023end}, a tracker needs to function in real time. An automated tracker can follow along a student as they are reading and when they are stuck at a difficult to pronounce word, it can prompt the word thus aiding the student. However, automated tracking is not without challenges. Children's reading, when they are learning, is especially difficult to track owing to the disfluencies present. There can be a lot of false starts, word repetitions and word skipping involved. 

Traditional modeling of a reading tracker has used a cascade of an ASR model and a rule based tracking algorithm \cite{rasmussen2011evaluating}. Li et al. \cite{li2012evaluating} further improve this method by taking into account the real time nature of tracking, but their method is also dependant on an ASR model. For a scenario where data is scarce, training an ASR model can be challenging. An example of this is the Reading Races dataset that we experiment with. In such cases, using an off-the-shelf pretrained ASR model can lead to hallucinations. Another problem is the time delay between the occurrence of acoustic evidence and the prediction \cite{plantinga2019towards}.


In this work, we build a fully end-to-end (E2E) speech tracker using a pointer network \cite{vinyals2015pointer}. This formulation is completely ASR free and our tracker learns an attention map over the text being read conditioned on the streaming speech. This attention map is learnt explicitly using ground truth alignments that we obtain from a forced aligner. Using this formulation, the time lag between acoustic evidence and the prediction is reduced to a significant extent as we directly predict a pointer position at each time step without having to predict the actual word. Another advantage of this approach is that we can directly get the alignment by reading the attention maps without needing to run a separate alignment algorithm. This way, we avoid the cascading effects of ASR errors.

We experiment with three forced alignment models to generate the ground truth to train the tracker, attention-based encoder-decoder ASR model (AED) \cite{chan2016listen}, a CTC-based ASR model \cite{graves2006connectionist} and the classical GMM-HMM based ASR model \cite{mcauliffe2017montreal}. We note that the advantage of using the AED model is that we naturally get soft alignments as training targets for the tracker which can be useful for knowledge distillation.

We provide results on one adult and two children speech datasets. For the adult voice dataset, we use TIMIT data \cite{garofolo1993timit}. For the children voice dataset, we use the CMU Kids \cite{eskenazi1997cmukids} data and the Reading Races dataset \cite{council2019improving}. We observe better tracking accuracy on TIMIT with the best tracking accuracy of 87.8\%. On CMU Kids and Reading Races, we report the best tracking accuracy of 77.1\% and 65.3\% respectively. We also provide qualitative results on the children's dataset showing how some disfluencies are is handled by the pointer network.

\section{Model Overview}
\label{sec:model_over}

\begin{figure}
    \hfill
    \centering
    \centerline{\includegraphics[width=\columnwidth]{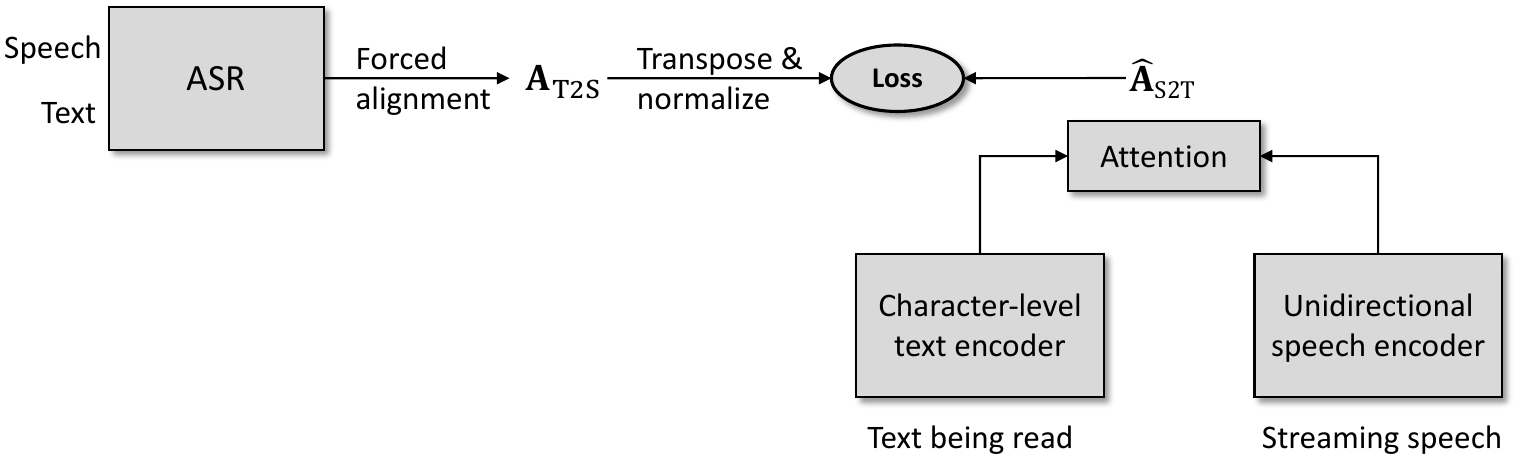}}
\caption{The ASR is used to generate the alignment between the speech and text, $\textbf{A}_{T2S} \in \mathbb{R}^{m \times n}$ where $m$ and $n$ are the number of text and speech tokens respectively. The tracker learns an attention over the text encoder output using the streaming speech to predict the alignment $\hat{\textbf{A}}_{S2T} \in \mathbb{R}^{n \times m}$ which is learnt using $\textbf{A}_{T2S} \in \mathbb{R}^{m \times n}$ as the supervision.} 
\label{fig:model_over}
\end{figure}

The overall pipeline for building a real time tracker comprises of two steps. In the first step, we generate forced alignments for the training data using an ASR-based model. These alignments are then used as the ground truth supervision for the second step which is to train a pointer network based tracker.

\subsection{Forced alignment}
\label{subsec:forced_aln}
A forced aligner is an ASR model which predicts the best possible alignment between a text input and the corresponding speech. We denote the alignment generated by the forced aligner as a matrix $\textbf{A}_{T2S} \in \mathbb{R}^{m \times n}$ where $m$ and $n$ are the number of text and speech tokens respectively. Each row of $\textbf{A}_{T2S}$ is the alignment of a text token with all the speech frames. Depending on the type of ASR model used, this alignment can be soft (a probability distribution) or hard (a one/multi-hot vector). We explore three ASR architectures for performing forced alignment.\\
\noindent \textbf{Attention-based encoder decoder (AED)}: This model is based on the LAS framework \cite{chan2016listen}. Here, speech is encoded using a bidirectional speech encoder and the decoder is a unidirectional LSTM which decodes the text one character at a time while implicitly learning an alignment between speech and text through the attention layer. The alignment matrix $\textbf{A}_{T2S}$ is obtained from the attention layer and is a soft alignment, i.e. for each character in the text, we obtain a probability distribution over the sequence of speech frames, denoting the alignment. We can also convert this soft alignment into a hard alignment easily by converting $\textbf{A}_{T2S}$ into a multi-hot vector based on an alignment-weight threshold.\\
\noindent \textbf{CTC-based ASR}: For this, we train an ASR model with the CTC criterion \cite{graves2006connectionist}. Once the model is trained, we follow Kurzinger et al. \cite{kurzinger2020ctc} to obtain the trellis matrix which is the probability of the characters aligned at each time step. Using this trellis, we can estimate the most likely CTC path for the given speech-text pair by backtracking. This gives us the desired alignment, $\textbf{A}_{T2S}$ which is a hard alignment.\\
\noindent \textbf{GMM-HMM based ASR}: We use the Montreal Forced Aligner (MFA) \cite{mcauliffe2017montreal} for this. We train acoustic models using the pronunciation dictionary provided for the domain specific datasets for forced alignment. MFA gives hard alignments by default at the word and phone level. We could also get soft alignments at the phonetic or grapheme level by extracting the $\gamma$ probabilities in the HMM model. For this work, we limit ourselves to the default hard alignments from MFA.

\subsection{Pointer network based tracker}
\label{subsec:pointer_net}
Pointer networks were introduced in Vinyals et al. \cite{vinyals2015pointer} for tackling various combinatorial problems with deep learning models. The original formulation of pointer networks is autoregressive, where the decoder points to a certain position in the encoder sequence and this position is then added to the decoder output. Our tracker application does not need the autoregressive formulation as our model is not generative.

Our pointer network consists of a character-level text encoder, a unidirectional LSTM-based speech encoder and an additive attention layer. Let the output of the text encoder be a sequence of character embeddings $(g_1, g_2, ..., g_m)$ and that of the speech encoder be a sequence of speech frames $(h_1, h_2, ..., h_n)$. Given these two sequences, we want to get an alignment estimate, $\hat{\textbf{A}}_{S2T} \in \mathbb{R}^{n \times m}$ where each row corresponds to a speech frame alignment with all characters in the text. We estimate $\hat{\textbf{A}}_{S2T}$ using the attention layer as follows,
\begin{align*}
    \begin{split}
        x_j^i &= \textbf{v}^\text{T} \text{tanh}(\textbf{W}_1 g_i + \textbf{W}_2 h_j) \\
        \textbf{a}_j &= \text{softmax}(\textbf{x}_j) \\
        \hat{\textbf{A}}_{S2T} &= \text{concat}([\textbf{a}_1, \textbf{a}_2, ..., \textbf{a}_n])
    \end{split}
\end{align*}

Here, $\textbf{v}$, $\textbf{W}_1$ and $\textbf{W}_2$ are learnable parameters. The alignment of the $j^{\text{th}}$ speech frame with all the characters in the text is denoted by the probability distribution $\textbf{a}_j$. 

At inference, we can compute alignments as we get speech frames, $h_j$ in real time using the unidirectional LSTM.\\

\noindent \textbf{Training}: To train the tracker, we obtain supervision from the alignments, $\textbf{A}_{T2S}$ generated from the forced aligner. We compute the ground truth, $\textbf{A}_{S2T}$ as follows,
\begin{align*}
    \begin{split}
        \textbf{A}_{S2T} = \text{L1-normalize}(\textbf{A}_{T2S}^\text{T})
    \end{split}
\end{align*}
We transpose $\textbf{A}_{T2S}$ and normalize it row-wise so that we obtain a probability distribution for every speech frame. If $\textbf{A}_{S2T}$ is a hard alignment, we compute,
\begin{align*}
    \begin{split}
        L_{hard} = \frac{1}{|\mathbb{B}|}\sum_{b \in \mathbb{B}}\frac{1}{N_b}\sum_{i=1}^{N_b}\text{CrossEntropy}(\hat{\textbf{A}}_{S2T}[i], \textbf{A}_{S2T}[i])
    \end{split}
\end{align*}
where the cross entropy is computed for the alignment of every speech frame, i.e. every row of the alignment matrices across the batch $\mathbb{B}$. When $\textbf{A}_{T2S}$ is a soft alignment, we compute,
\begin{align*}
    \begin{split}
        L_{soft} = \frac{1}{|\mathbb{B}|}\sum_{b \in \mathbb{B}}\frac{1}{N_b}\sum_{i=1}^{N_b}\text{KLDivergence}(\hat{\textbf{A}}_{S2T}[i], \textbf{A}_{S2T}[i])
    \end{split}
\end{align*}
$L_{soft}$ can also be computed as a cross entropy loss, but we follow previous work by Hinton et al. \cite{hinton2015distilling} and use KL-Divergence by treating the soft target as a knowledge distillation target. The model overview is shown in figure \ref{fig:model_over}.

\section{Experiments}
\label{sec:experiments}
We train two deep learning based E2E ASR systems to build the AED and CTC forced aligners. For AED, we follow the design choices of Chan et al. \cite{chan2016listen}. We use one BiLSTM layer to encode sequences of 80 dimensional log mel-filterbank features, followed by two pyramid BiLSTM layers which downsample the sequence length by a factor of 4. Finally, we add the last BiLSTM layer to produce the final acoustic representations. The decoder is a two layer LSTM and the attention mechanism is content-based and additive.

The CTC model has the same speech encoder architecture as AED. We train the CTC and AED models on Librispeech \cite{panayotov2015librispeech} followed by an adaptation on the downstream datasets.

For the tracker, the text encoder is a two layer BiLSTM and the speech encoder follows the same architecture as the encoder of AED except that it is unidirectional. The tracker is pre-trained on Librispeech and adapted on downstream data.

\subsection{Evaluation}
\label{subsec:eval}
\noindent \textbf{Forced alignment}: To evaluate the performance of forced aligners, we use precision, recall and jaccard similarity. For a given word, let $t_1$ and $t_2$ be the ground truth start and end times and $\hat{t}_1$ and $\hat{t}_2$ be the predicted start and end times respectively. Then, we define the evaluation metrics as follows,
\begin{align*}
    \begin{split}
        \text{intersection} &= \text{max}(\text{min}(t_2, \hat{t}_2) - \text{max}(t_1, \hat{t}_1), 0) \\
        \text{union} &= t_2 - t_1 + \hat{t}_2 - \hat{t}_1 - \text{intersection} \\
        \text{jaccard (\textbf{Ja})} &= \frac{\text{intersection}}{\text{union}} \\
        \text{precision (\textbf{Pr})} &= \frac{\text{intersection}}{\hat{t}_2 - \hat{t}_1} \\
        \text{recall (\textbf{Re})} &= \frac{\text{intersection}}{t_2 - t_1}
    \end{split}
\end{align*}

\begin{figure}
    \hfill
    \centering
    \centerline{\includegraphics[width=0.7\columnwidth]{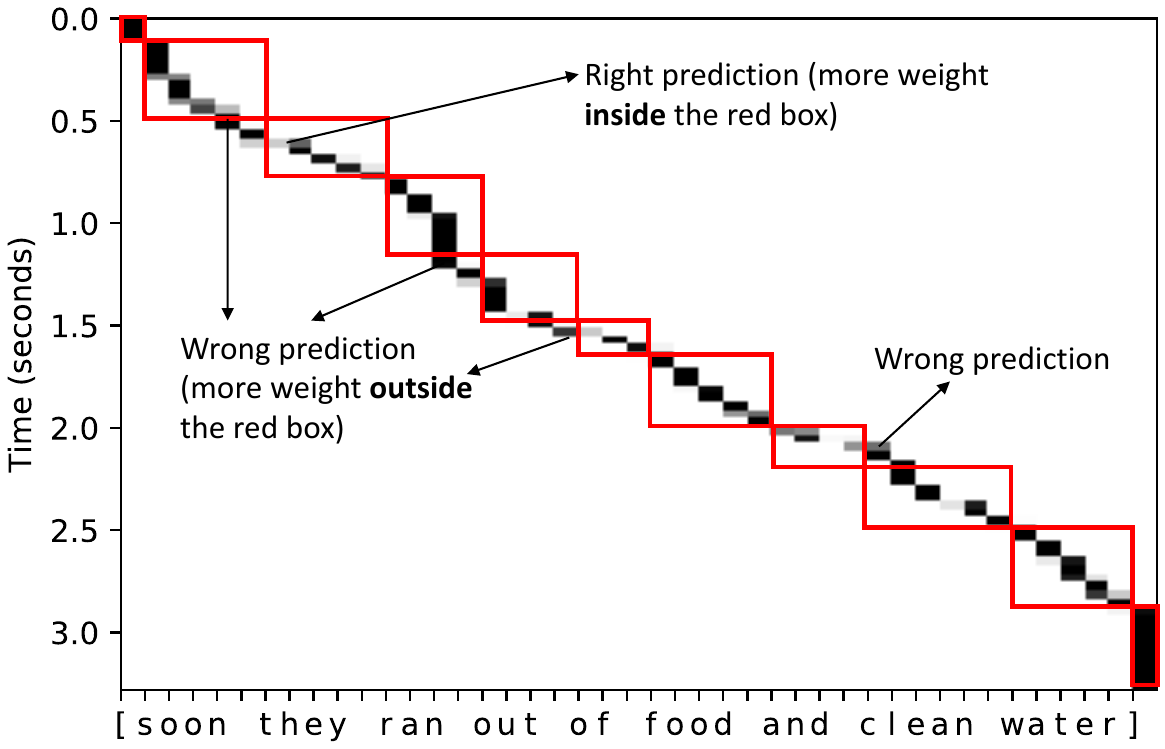}}
\caption{The red boxes are the ground truth alignments of words against time. The pixels represent the tracker prediction. The height of each pixel is a 40ms frame. For a correct word prediction by a frame, the total weight inside the red box should be greater than that outside. For all frames, we can count the number of correctly/incorrectly predicted words.}
\label{fig:eval_tracker}
\end{figure}

\noindent \textbf{Tracking}: To evaluate the tracking performance of the pointer network, we use the predicted alignment, $\hat{\textbf{A}}_{S2T}$. Each row of this matrix represents the alignment of a 40ms speech frame with all characters in the text. For each frame, we compute the score for every word in the text by adding the character weights for that word in the corresponding row. The word with the highest score is then predicted. A detailed illustration of this is shown in figure \ref{fig:eval_tracker}. To make the prediction for a speech frame $i$ more deterministic, we make the output distribution sharper using the following operation with $\tau = 0.1$,
\begin{align*}
    \begin{split}
        \texttt{sharp}(\hat{\textbf{A}}_{S2T}[i]) = \frac{(\hat{\textbf{A}}_{S2T}[i])^{\frac{1}{\tau}}}{||(\hat{\textbf{A}}_{S2T}[i])^{\frac{1}{\tau}}||_1}
    \end{split}
\end{align*}


\begin{table}
    \centering
    \resizebox{\columnwidth}{!}{
    \begin{tabular}{@{}lccccccccccc@{}}\toprule
         & \multicolumn{3}{c}{TIMIT} && \multicolumn{3}{c}{CMUK} && \multicolumn{3}{c}{READR}\\
         Ground truth & \multicolumn{3}{c}{Manual} && \multicolumn{3}{c}{Sphinx II \cite{huang1993sphinx}} &&  \multicolumn{3}{c}{Manual (10 samples)}\\
         \midrule
         & Pr & Re & Ja && Pr & Re & Ja && Pr & Re & Ja \\
         \midrule
         AED-aligner (ours) & 84.57 & 89.65 & 76.16 && 78.87 & \textbf{82.98} & 68.10 && \textbf{80.91} & \textbf{74.94} & \textbf{73.15} \\
         CTC-aligner & 61.93 & 69.58 & 50.29 && 50.80 & 58.26 & 37.49 && 39.33 & 61.34 & 31.45 \\
         MFA (flat start) & \textbf{91.19} & \textbf{90.96} & \textbf{83.56} && \textbf{84.72} & 75.30 & \textbf{68.48} && 21.57 & 37.00 & 17.48 \\
         MFA (adapted) & 48.66 & 49.48 & 34.91 && 71.03 & 67.51 & 56.78 && 16.98 & 24.17 & 12.72 \\
         \bottomrule
    \end{tabular}}
    \caption{Forced alignment results on TIMIT, CMUK and READR. The ground truth for TIMIT is manually annotated, for CMUK, we use the provided Sphinx II annotation as ground truth. For READR, we manually annotate 10 random examples from the test set for evaluation.}
    \label{tab:results_fa}
\end{table}

\subsection{Datasets}
\label{subsec:data}
We use three datasets for evaluation.\\
\noindent \textbf{TIMIT \cite{garofolo1993timit}}: This is a 5-hour dataset of adult voice recordings. We use the standard train-test split. This data provides time aligned transcriptions which act the ground truth.\\
\noindent \textbf{CMU Kids (CMUK) \cite{eskenazi1997cmukids}}: This is a 9-hour corpus of children read speech. The children's age vary between 6 to 11 years. This data provides time aligned transcriptions from Sphinx II \cite{huang1993sphinx}. For disfluencies, we are provided with a phoneme level transcription. We convert these into word-level transcription by using dynamic time warping to align orthographic and phonemic transcriptions of words.\\
\noindent \textbf{Reading Races (READR) \cite{council2019improving}}: This is a 15-hour corpus of children read speech with each data instance being a minute long. This is a more challenging dataset with participants being in the age group of 5 to 8 years with reading difficulties.

\subsection{Results}
\label{subsec:results}
\noindent \textbf{Forced alignment}: The output of forced alignment acts as the training signal for our pointer-network based tracker. We compare forced alignment performance of three ASR models: AED, CTC and GMM-HMM (MFA for montreal forced aligner \cite{mcauliffe2017montreal}). The results are shown in table \ref{tab:results_fa}.

We note that MFA performs best compared to the other two on TIMIT and CMUK. However, this required training an acoustic model using the provided pronunciation dictionary. AED and CTC based forced aligners are fully E2E and do not require a pronunciation dictionary to adapt their acoustic models. The advantage of this is evidenced by the results of MFA on READR which does not have a pronunciation dictionary of its own and thus MFA performs poorly. For READR, we manually time aligned 10 one-minute long random examples from the test set. We note that AED performs much better compared to CTC with an additional advantage that AED provides soft alignments between speech and text which can be used for teacher forcing to train the tracker.

\noindent \textbf{Tracking}: Table \ref{tab:results_tr} shows the tracking results. We use the evaluation procedure mentioned in section \ref{subsec:eval} reporting tracking accuracy and F1 score. We report tracking performance with 4 training signals but do not evaluate the CTC based training signal as it's forced alignment performance was not very good.

\begin{table}
    \centering
    \resizebox{\columnwidth}{!}{
    \begin{tabular}{@{}lcccccccc@{}}\toprule
          & \multicolumn{2}{c}{TIMIT} && \multicolumn{2}{c}{CMUK} && \multicolumn{2}{c}{READR}\\
          Ground truth ($\rightarrow$) & \multicolumn{2}{c}{Manual} && \multicolumn{2}{c}{Sphinx II \cite{huang1993sphinx}} && \multicolumn{2}{c}{AED}\\ 
         \midrule
         Training signal ($\downarrow$) & Acc & F1 && Acc & F1 && Acc & F1\\
         \midrule
         AED ($L_{hard}$) & 83.30 & 81.26 && 73.92 & 70.82 && \textbf{65.34} & \textbf{67.76}\\
         AED ($L_{soft}$) & 87.73 & 83.81 && \textbf{77.07} & \textbf{77.15} && 63.68 & 63.41 \\
         AED ($L_{hard}+L_{soft}$) & \textbf{87.82} & \textbf{83.85} && 77.06 & 76.13 && 64.45 & 67.12 \\
         MFA & 77.67 & 72.76 && 67.11 & 49.70 && 7.69 & 5.84\\
         \bottomrule
    \end{tabular}}
    \caption{Tracking results. We report the results of using the training signal from 4 different forced aligner configurations to train to tracker. The accuracy and F1 scores are measured against in the manner described in figure \ref{fig:eval_tracker}. For READR, we use the AED forced aligner output as the ground truth.}
    \label{tab:results_tr}
\end{table}

When using the AED variants for the training signal, $\textbf{A}_{S2T}$, we perform significantly better compared to the MFA based $\textbf{A}_{S2T}$ even though MFA gave better forced alignment results on TIMIT and CMUK. We observe that this is because MFA gives word-level time alignments while the AED model is trained to give character-level alignments. Hence, the tracker can also be trained at the character level. For a character-level model, even if a character prediction for a frame is out of bounds (outside the red box in figure \ref{fig:eval_tracker}), there is still a chance for recovery through its alignment with other characters in the word (more weight inside the red box in figure \ref{fig:eval_tracker}). There is no such recovery provision for a word-level model.

For the READR data, we note that using the hard training signal ($L_{hard}$) gives best performance whereas for CMUK and TIMIT, the soft signal ($L_{soft}$) helps more. We hypothesize that as READR has very long audio inputs (1 minute on average), having a precise alignment as a training signal better facilitates the model in finding the exact location. Also note that the MFA training signal performs poorly on READR due to the MFA forced alignments themselves being very poor in table \ref{tab:results_fa}.

\begin{table}
    \centering
    \resizebox{0.8\columnwidth}{!}{
    \begin{tabular}{@{}lccccc@{}}\toprule
         & \multicolumn{2}{c}{Manual annotation} && \multicolumn{2}{c}{AED annotation} \\
         \midrule
         Training signal & Acc & F1 && Acc & F1 \\
         \midrule
         AED ($L_{hard}$) & \textbf{69.34} & \textbf{71.99} && 64.08 & 67.25 \\
         \bottomrule 
    \end{tabular}}
    \caption{Tracker result on 10 random examples from READR, comparing manual alignments with AED alignments.}
    \label{tab:human_ann}
\end{table}


For READR, we also evaluate the tracker against the 10 manually aligned examples. We compare this with the performance we got when evaluating against the force aligned ground truth (see table \ref{tab:human_ann}). Note that we see better performance when evaluating against the manual annotations which shows that our tracker follows closer with human alignment compared to automated alignments.

\begin{figure}
    \hfill
    \centering
    \centerline{\includegraphics[width=0.75\columnwidth]{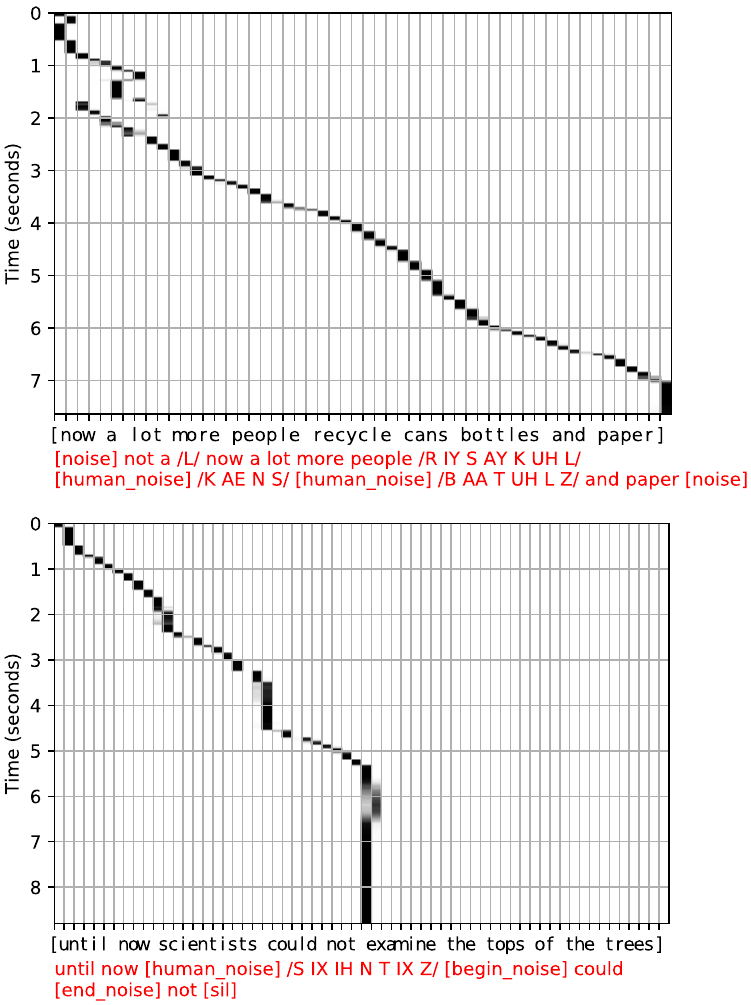}}
\caption{Tracking output of the pointer network for two examples from CMUK. X-axis is the actual sentence being read and the text in red is the transcript. We see that word repetition (top) and skipping (bottom) is effectively captured.}
\label{fig:qual_eval}
\end{figure}

Finally, we show how the tracker behaves in the presence of disfluencies in figure \ref{fig:qual_eval}. The top plot shows that in the case of repetition or false start, the tracker is able to effectively realign itself and continue the monotonic trajectory. Similarly, the tracker is able to detect the stopping point in a partially read sentence and ignore the unread part (see bottom of figure \ref{fig:qual_eval}). Thus, we see qualitatively that the tracker can give meaningful alignments for disfluent children's speech.

\section{Conclusion}
In this work, we build a real time reading tracker using pointer network. Our proposed method does not require manual annotation and relies on forced alignment to generate the training signal to train the tracker. We explore different forced alignment strategies to generate the training signal and note that AED based forced alignment works best to train the tracker.

\vfill\pagebreak

\bibliographystyle{IEEEbib}
\bibliography{strings,refs}

\end{document}